\documentclass[]{jkas} 

\def\year{2026} 
\def\volume{---} 
\def\issue{---} 
\def\beginpage{1} 
\def\received{---} 
\def\accepted{---} 
\def\published{---} 
\date{Received \received; Accepted \accepted; Published \published}

\usepackage{flushend} 
\newcommand\ion[2]{{#1}\,\textsc{#2}} 

\title{Comparing the Nebular and Stellar Dust Attenuation Curves in Nearby Star-Forming Galaxies}

\author[1,$\star$]{Jong Chul Lee}{0000-0003-0283-8352}
\author[2]{Hyunjin Shim}{0000-0002-4179-2628}
\author[3]{Hyunmi Song}{0000-0002-4362-4070}
\author[4,5,6]{Ho Seong Hwang}{0000-0003-3428-7612}
\author[1]{Woong-Seob Jeong}{0000-0002-2770-808X}
\author[1]{Jong-Ho Shinn}{0000-0001-7967-6473}

\affil[1]{Korea Astronomy and Space Science Institute (KASI), 776 Daedeokdae-ro, Yuseong-gu, Daejeon 34055, Republic of Korea}
\affil[2]{Department of Earth Science Education, Kyungpook National University, 80 Daehak-ro, Buk-gu, Daegu 41566, Republic of Korea}
\affil[3]{Department of Astronomy and Space Science, Chungnam National University, 99 Daehak-ro, Yuseong-gu, Daejeon 34134, Republic of Korea}
\affil[4]{Astronomy Program, Department of Physics and Astronomy, Seoul National University, 1 Gwanak-ro, Gwanak-gu, Seoul 08826, Republic of Korea}
\affil[5]{SNU Astronomy Research Center, Seoul National University, 1 Gwanak-ro, Gwanak-gu, Seoul 08826, Republic of Korea}
\affil[6]{Institute for Data Innovation in Science, Seoul National University, Seoul 08826, Republic of Korea}

\def\corrauthor{%
J. C. Lee, \email{jclee@kasi.re.kr}
}

\def\runningauthor{%
Lee et al.
}

\def\runningtitle{%
Nebular versus Stellar Attenuation Curves
}

\def\keywords{%
dust, extinction --- galaxies: ISM --- galaxies: star formation --- catalogues   
}

\def\abstracttext{%
Differences in the spatial distributions of stars and ionized gas relative to dust in galaxies imply that stellar and nebular dust attenuation curves need not be identical.
Although stellar attenuation curves have been extensively studied, nebular attenuation curves remain less explored. 
In this work, we investigate nebular attenuation curves using Balmer line ratios for a sample of 90,958 nearby star-forming galaxies from the Sloan Digital Sky Survey with well-constrained stellar attenuation curves.
In the H$\alpha$/H$\beta$ versus H$\alpha$/H$\gamma$ diagram, the nebular attenuation curves are, on average, significantly shallower than typical stellar attenuation curves, which can be qualitatively reproduced by source-dust geometric effects.
The nebular attenuation curve slopes show weak positive correlations with the stellar attenuation curve slopes and weak dependences on galaxy properties that broadly follow those of the stellar attenuation curve slopes.
These results suggest that nebular attenuation curves are distinct from stellar attenuation curves, while remaining consistent with a connection between them.
}

\begin{document}
\jkashead 


\section{Introduction}\label{sec:intro}

Dust in galaxies, although comprising only a minor fraction of the baryonic mass, plays an important role in the processes of galaxy evolution.
Dust grains are primarily formed in evolved-star winds and supernova ejecta, grow through accretion in the interstellar medium, and are destroyed by shocks and intense radiation fields \citep[e.g.,][]{aoy17,gal21,dub24}.
By facilitating gas cooling and shielding molecular hydrogen from dissociating hard radiation, interstellar dust can enhance star formation activity \citep[e.g.,][]{hir02,gne09}. 
Dust also serves as a reservoir of cosmic metals, which are essential for the formation of planets and biological molecules in the universe \citep[e.g.,][]{obe21,kra22}.

From an observational perspective, one of the most significant impacts of dust is modifying the spectra of galaxies (see \citealt{gal18} for a review and references therein).
By preferentially absorbing and scattering ultraviolet (UV) and blue optical photons and re-emitting the absorbed energy in the infrared (IR) regime as thermal radiation, dust redistributes galaxy light across wavelengths, leading to dimming and reddening of the observed spectrum.
As a result, dust complicates the interpretation of galaxy properties such as star formation rate (SFR), stellar mass, age, and metallicity. 
A proper understanding of the dust effect is crucial for accurately recovering the intrinsic galaxy properties and for constructing a consistent picture of galaxy evolution.

Dust extinction and attenuation curves are functions that quantify the wavelength-dependent modification of galaxy spectra caused by dust.
Extinction indicates absorption and scattering out of the line of sight and is directly linked to the underlying physics of dust, which is governed by the grain size distribution and chemical composition \citep[e.g.,][]{wei01,hir20}.
Attenuation encompasses scattering back into the sightline and the contribution from unobscured light sources, reflecting the complexity of the source-dust geometry in galaxies.
Conventionally, the term attenuation is used for spatially unresolved studies.

In most studies, dust attenuation curves refer to stellar attenuation curves that describe the attenuation of stellar continuum light.
They have been extensively explored mainly through spectral energy distribution (SED) fitting techniques and show substantial galaxy-to-galaxy variation \citep[e.g.,][]{nol09,kri13,sal19,bat20,rod26,shi26}. 
A well-known behavior is that more opaque galaxies have attenuation curves with weaker wavelength dependence (i.e., shallower/flatter/grayer curves), as expected from radiative transfer and geometric effects \citep[e.g.,][]{bae01,che13,tra20}.
Various empirical trends between their shape and other physical parameters have also been reported \citep[e.g.,][]{wid11,bat16,sal18,shi20,mar25,wij26}.
A comprehensive review can be found in \citet{sal20}.

Compared with stellar attenuation curves, nebular attenuation curves, which characterize the attenuation of ionized gas emission, are poorly understood.
Stellar attenuation curves are inherently entangled with uncertainties in stellar population properties and star formation histories \citep[e.g.,][]{shi11,lej17,qin22}, whereas nebular attenuation curves can be more robustly constrained by comparing observed hydrogen recombination line ratios with their theoretical predictions (but see also \citealt{ji23}).
Nevertheless, nebular attenuation curve studies have been limited by the difficulty of obtaining homogeneous emission-line measurements spanning a wide wavelength range for large samples of galaxies.
Due to the lack of a generally accepted nebular attenuation curve, the stellar extinction/attenuation curves of \citet{car89}, \citet{odo94}, and \citet{cal00} have been commonly adopted instead, for example, when deriving dust-corrected H$\alpha$ luminosities as SFR indicators \citep[e.g.,][]{ken09,lee13,red15,qin19,val20}.
The amounts of stellar and nebular attenuation are clearly correlated, but the stellar-to-nebular attenuation ratio varies significantly, ranging from 0.44 to 1.0 \citep[e.g.,][]{pug16,shi20,lee25}. 
This suggests that the common assumption that nebular attenuation curves are similar to stellar attenuation curves needs to be reconsidered, emphasizing the importance of applying appropriate nebular attenuation curves for dust corrections \citep[e.g.,][]{lor26,pah26,red26b}.

\citet{bau95} presented one of the pioneering studies by deriving a nebular extinction curve in the Orion nebula using the \citet{car89} extinction curve with an additional wavelength-dependent term.
\citet{lee12} combined SDSS H$\alpha$ and H$\beta$ data with AKARI Br$\alpha$ measurements to provide observational evidence that nebular attenuation curves of IR-luminous galaxies may be shallower than the \citet{cal00} attenuation curve. 
\citet{red20} and \citet{rez21} stacked Keck spectra of high-$z$ galaxies and SDSS spectra of low-$z$ galaxies, respectively, to detect even high-order Balmer lines, finding that nebular attenuation curves are broadly consistent with typical stellar attenuation curves.
\citet{lin24} reported similar results based on MaNGA data.
\citet{pre22} and \citet{woz26} showed that the large scatter in the H$\alpha$/H$\beta$ versus Pa$\beta$/H$\alpha$ plane can be partially attributed to variations in the dust-covering fraction.
\citet{san25} and \citet{red26a} utilized JWST spectroscopy covering optical to near-IR wavelengths and revealed that nebular attenuation curves are significantly deviated from typical stellar attenuation curves, although they rely on a small number of high-$z$ galaxies (but see also \citealt{coo25}).
However, despite these efforts, direct comparisons between nebular and stellar attenuation curves are still lacking, and the relationship between nebular attenuation curve and other galaxy properties has yet to be explored in detail.

In this work, we aim to constrain the slopes of nebular attenuation curves by analyzing Balmer line ratios for the galaxy sample of \citet{sal18}, whose stellar attenuation curves are well characterized, and to systematically compare nebular and stellar attenuation curve slopes for the same galaxies.
We also investigate correlations between the nebular attenuation and other galaxy parameters, and examine whether the trends differ from those established for stellar attenuation.
The structure of this paper is as follows. 
Section 2 explains the datasets used and the sample selection. 
Section 3 presents the results of fitting the nebular attenuation curve slope and examining its dependence on various physical properties, including the stellar attenuation curve slope, with related discussions.
Our findings are summarized in Section 4. 
We follow the assumptions of \citet{sal18}, namely a \citet{cha03} initial mass function and a WMAP7 flat cosmology with $H_0$ = 70.4 km s$^{-1}$ Mpc$^{-1}$ and $\Omega_m$ = 0.272.

\section{Data and Sample}\label{sec:data}

\subsection{Catalogs}\label{sec:catalog}

\begin{figure*}[ht]
\centering
\includegraphics[width=\textwidth]{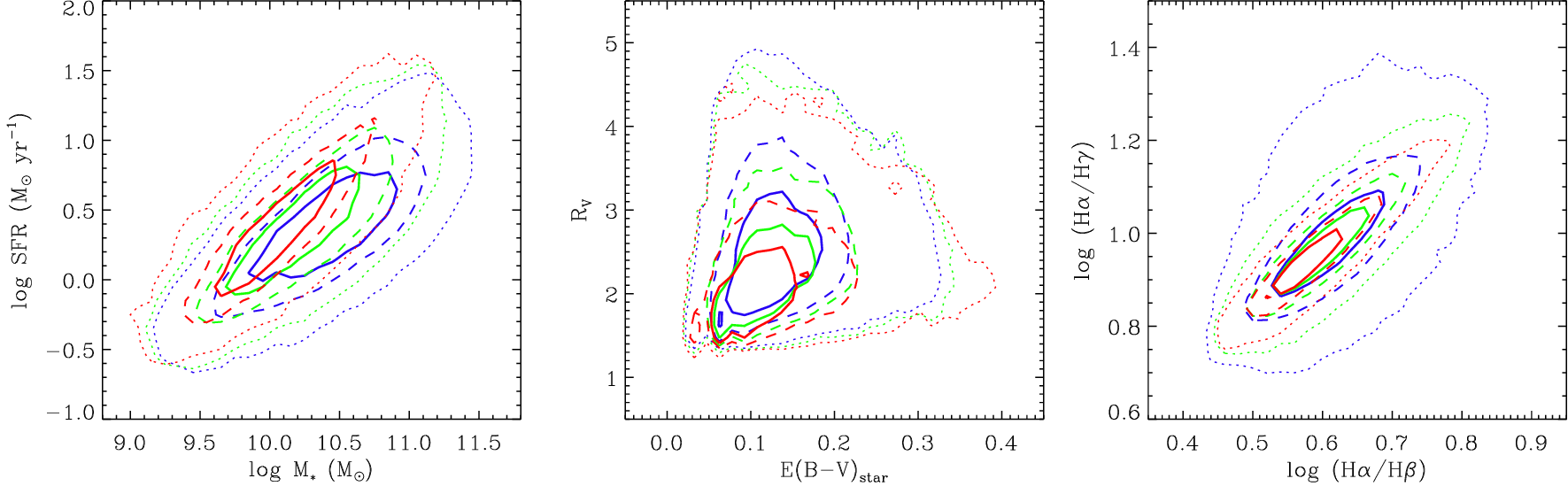}
\caption{Sample galaxy distributions in the stellar mass--star formation rate (left), $E(B-V)_{\rm star}$--$R_V$ (middle), and H$\alpha$/H$\beta$--H$\alpha$/H$\gamma$ line ratio (right) planes. The blue, green, and red lines denote the number density contours for the sample galaxies satisfying S/N$_{\rm line} >$ 1, 5, and 10, respectively. The solid, dashed, and dotted lines correspond to the 0.5$\sigma$ (38.3\%), 1$\sigma$ (68.3\%), and 2$\sigma$ (95.4\%) enclosures.\label{fig_sample}}
\end{figure*}

\citet{sal18} derived and released physical properties of nearby ($z < 0.3$) galaxies based on photometric data from the far-UV to the mid-IR (GALEX--SDSS--WISE Legacy Catalog; GSWLC\footnote{\url{https://salims.pages.iu.edu/gswlc/}}).
They used the stellar population synthesis models of \citet{bru03} and the Code Investigating GALaxy Emission (CIGALE; \citealt{boq19}) to perform SED fitting, and parameterized the stellar attenuation curve by modifying the \citet{cal00} law to allow its curve slope to vary.
From the catalog, we obtain the stellar mass ($M_*$), SFR, and stellar attenuation in the $B$ and $V$ bands ($A_B$, $A_V$).
We then calculate the stellar color excess, $E(B-V)_{\rm star}=A_B-A_V$, and the shape parameter of the stellar attenuation curve, $R_V=A_V/(A_B-A_V)$. 
Larger (smaller) values of $R_V$ mean shallower (steeper) stellar attenuation curves.

We retrieve optical emission-line fluxes, equivalent widths, and associated errors from the MPA–JHU catalog\footnote{\url{https://wwwmpa.mpa-garching.mpg.de/SDSS/DR7/}} \citep{bri04,tre04}, based on the fiber spectroscopic data of the SDSS.
The line fluxes were measured from Gaussian fits to the continuum-subtracted spectra and were corrected for foreground Galactic reddening with the \citet{sch98} dust map and the \citet{odo94} extinction curve.
The oxygen abundance 12 + log(O/H), estimated from ([\ion{O}{ii}]$\lambda3727+$[\ion{O}{iii}]$\lambda\lambda4959,5007$)/H$\beta$ \citep{tre04}, is used as a gas-phase metallicity indicator.

The $i$-band isophotal minor-to-major axis ratio ($b/a$) is also taken from the KIAS value-added galaxy catalog\footnote{\url{https://astro.kias.re.kr/vagc/dr7/}} \citep{cho10}, as a proxy for galaxy inclination.

\subsection{Sample Selection}\label{sec:select}

From the 659,229 galaxies in the GSWLC, we first select 386,049 galaxies that have detections in at least one of the far- or near-UV bands and in at least one of the 12 or 22 $\mu$m bands\footnote{We use the GALEX and WISE detection flags in the GSWLC, which are based on the 3$\sigma$ and 2$\sigma$ thresholds, respectively.}.
Among them, 379,171 galaxies whose SED fitting results are reliable ($0<A_V<A_B$) are adopted.
Although \citet{sal18} considered only GALEX medium-deep survey data, we utilize the all-sky data regardless of depth. 
Limiting the analysis to medium-deep galaxies reduces the sample size by about 40\%, but our conclusions remain unchanged.

We then cross-match the GSWLC galaxies with the MPA-JHU catalog via SDSS identification numbers, yielding 378,762 galaxies.
The GSWLC photometry captures the galaxy-integrated light, whereas the SDSS spectroscopy can be dominated by the central region because the spectra are obtained through fixed diameter fibers.
To mitigate aperture-mismatch effects, our sample is restricted to 337,219 galaxies at $z > 0.04$, following the suggestion of \citet{kew05}.
Considering that the redshift cut may still be insufficient for the most massive galaxies, we find broadly consistent results even when adopting a higher redshift cut.
For example, in our primary sample (defined in the following paragraph), the average nebular attenuation curve slopes are $1.337\pm0.002$ for 90,958 galaxies with $z > 0.04$ and $1.348\pm0.005$ for 25,469 galaxies with $z > 0.1$.
We classify the primary ionizing sources in galaxies using the standard Baldwin–Phillips–Terlevich diagram of [\ion{O}{iii}]$\lambda5007$/H$\beta$ versus [\ion{N}{ii}]$\lambda6584$/H$\alpha$ \citep{bal81} and focus on 180,572 star-forming galaxies that lie below the pure star formation line of \citet{kau03}.
Of these, we retain 174,467 galaxies with H$\alpha$ equivalent width $>$ 3 \AA\ to remove residual contamination from passive galaxies \citep{cid10}.

To investigate the nebular attenuation curve using Balmer line ratios, we select 173,602 galaxies with measured H$\gamma$ fluxes in addition to H$\alpha$ and H$\beta$.
Although H$\delta$ and higher-order Balmer lines would extend the wavelength baseline, these weak lines are not considered in this study to avoid a significant reduction in sample size and increased measurement scatter in Balmer line ratio diagrams.
Defining S/N$_{\rm line}$ as the minimum signal-to-noise ratio in the five emission lines (H$\alpha$, H$\beta$, H$\gamma$, [\ion{O}{iii}]$\lambda5007$, and [\ion{N}{ii}]$\lambda6584$), the numbers of galaxies satisfying S/N$_{\rm line} >$ 1, 5, and 10 are 166,305, 90,958 and 40,652, respectively, corresponding to 95.8\%, 52.4\% and 23.4\% of the total sample.
The three samples are introduced to assess the completeness and potential selection biases associated with different S/N$_{\rm line}$ thresholds. 
Of these, the S/N$_{\rm line} >$ 5 sample is the primary sample. 
Unless otherwise stated, the following analyses are based on this sample.

Figure~\ref{fig_sample} exhibits the distributions of key properties of the sample galaxies and their dependence on different S/N$_{\rm line}$ cuts.
In the $M_*$--SFR panel (left), the sample galaxies form a linear relation, the so-called star-forming main sequence \citep[e.g.,][]{elb07,whi12}.
At higher S/N$_{\rm line}$ cuts, the sequence not only becomes tighter but also shifts toward higher specific star formation rates (sSFR; SFR/$M_*$).
This is mainly driven by the preferential selection of less massive galaxies rather than high-SFR galaxies.
Its slope also changes from below unity in the low-S/N$_{\rm line}$ sample to nearly unity in the high-S/N$_{\rm line}$ sample because the fraction of high-sSFR galaxies is lower in more massive galaxies.
The $E(B-V)_{\rm star}$--$R_V$ (middle) panel shows that on average high-S/N$_{\rm line}$ galaxies are less dusty and have steeper stellar attenuation curves than low-S/N$_{\rm line}$ galaxies.
A positive relation between $E(B-V)_{\rm star}$ and $R_V$ is clearly seen along the thick red/green contour, reaffirming the result of \citet{sal18} that dustier galaxies tend to have shallower stellar attenuation curves.
In the Balmer line ratio (right) panel, H$\alpha$/H$\gamma$ ratios are strongly correlated with H$\alpha$/H$\beta$ ratios.
As the S/N$_{\rm line}$ cut increases, outliers are eliminated and the locus appears to shift slightly toward the less dusty regime.
No other substantial changes are found, implying that the selection bias associated with the S/N$_{\rm line}$ threshold has little impact on our analysis of nebular attenuation curves.

\section{Results and Discussion}\label{sec:result}

\subsection{Constraining Nebular Attenuation Curve}\label{sec:slope}

\begin{figure}[!t]
\centering
\includegraphics[width=\columnwidth]{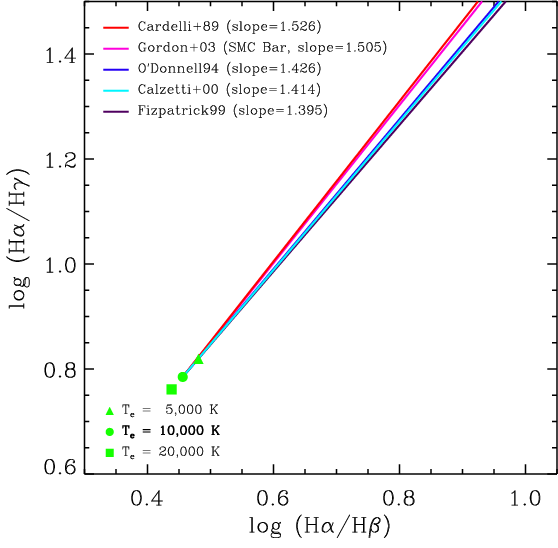}
\caption{Relation between H$\alpha$/H$\beta$ and H$\alpha$/H$\gamma$ line ratios, predicted by widely used dust extinction and attenuation curves: \citet[red]{car89}, \citet[magenta]{gor03}, \citet[blue]{odo94}, \citet[cyan]{cal00}, and \citet[purple]{fit99}. The green symbols indicate the intrinsic line ratios based on the Case B recombination. The triangle, circle, and square correspond to electron temperatures of $T_e$ = 0.5, 1.0, 2.0$\times$10$^4$ K, respectively, assuming an electron density of $n_e$ = 100 cm$^{-3}$. The line ratio relations are plotted starting from the commonly adopted case of $T_e$ = 10$^4$ K.\label{fig_model}}
\end{figure}

Figure~\ref{fig_model} demonstrates the predicted relation between log(H$\alpha$/H$\beta$) and log(H$\alpha$/H$\gamma$) for star-forming galaxies, assuming that they are dominated by \ion{H}{ii} regions, with the same intrinsic line ratios, and that their nebular lines are affected by dust according to a specific attenuation curve ($k_{\lambda}$). 
Under these conditions, star-forming galaxies move away from the intrinsic position along a straight line toward larger line ratios as dust attenuation increases.
The slope of the star-forming galaxy sequence in the log(H$\alpha$/H$\beta$) versus log(H$\alpha$/H$\gamma$) plane is directly connected to the slope of the nebular attenuation curve in selective form, $(k_{\rm H\gamma}-k_{\rm H\alpha})/(k_{\rm H\beta}-k_{\rm H\alpha})$.
Hereafter, we simply refer to this as the nebular attenuation curve slope.
Given the lack of commonly adopted nebular attenuation curves, we present the slopes inferred from stellar attenuation curves \citep[][]{car89,odo94,fit99,cal00,gor03}.
If the nebular lines are dust-attenuated following these typical attenuation curves, the slopes are expected to be about 1.40--1.53.
The hydrogen recombination line ratios depend weakly on electron temperature $T_e$ but change little with electron density $n_e$ \citep{dop03}.
We adopt H$\alpha$/H$\beta$ = 3.026, 2.857, 2.743 and H$\alpha$/H$\gamma$ = 6.595, 6.092, 5.770 for Case B recombination at $T_e=$ 0.5, 1.0, and 2.0$\times10^4$ K (with $n_e$ = 100 cm$^{-3}$), respectively, using the \texttt{PyNeb} package \citep{lur15}.
Interestingly, the slopes connecting the intrinsic positions are $\sim$1.36, roughly matching the average slope of the nebular attenuation curves ($\sim$1.34; see the following paragraphs). 
Therefore, changing the adopted intrinsic position has minimal influence on deriving the nebular attenuation curve slope indicators (see Section~\ref{sec:depend}). 
The similarity between the intrinsic and average slopes implies that the observed sequence may be affected by the intrinsic line ratio variation. 
However, the observed H$\alpha$/H$\beta$ (H$\alpha$/H$\gamma$) ratios span a 95\% range of 0.277 (0.406) dex, whereas the maximum variation in the corresponding intrinsic line ratios is only 0.043 (0.058) dex. 
Since typical star-forming galaxies are likely to cover a narrower range of physical conditions than these extreme cases, their intrinsic line ratios should vary even less. 
We thus expect the degeneracy introduced by intrinsic line ratio variations to cause only small changes in the derived nebular attenuation curve slopes.


\begin{figure*}[htbp]
\centering
\includegraphics[width=\textwidth]{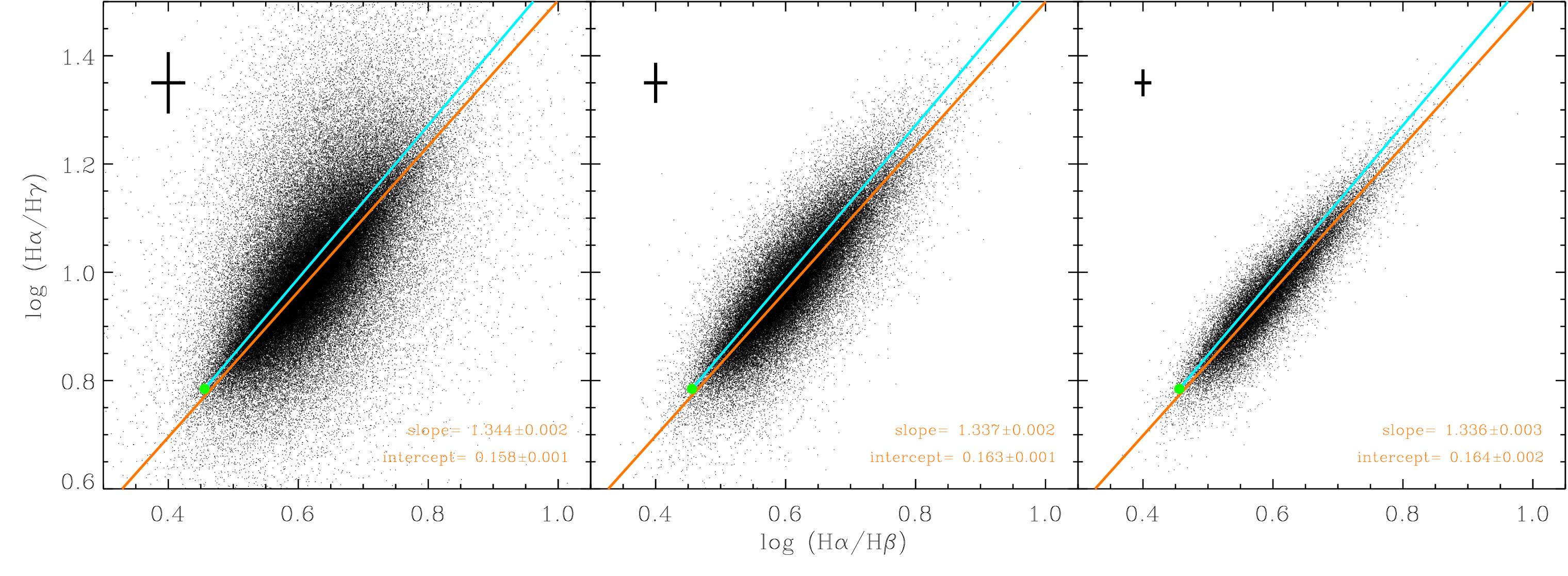}
\caption{H$\alpha$/H$\beta$ versus H$\alpha$/H$\gamma$ line ratio diagrams for the sample galaxies with S/N$_{\rm line} >$ 1 (left), 5 (middle), and 10 (right). In each panel, the green circle represents the location of dust-free star-forming galaxies assuming $T_e = 10^4$ K and the cyan line denotes the \citet{cal00} curve. The error bar in the upper-left corner indicates the median values of the line ratio uncertainties. The orange line shows the best-fit linear relation for the sample galaxies, with the fitted slope and intercept given in the lower-right corner, derived from an error-weighted orthogonal distance regression.\label{fig_slope}}
\end{figure*}

Figure~\ref{fig_slope} shows the observed Balmer line ratios of the sample galaxies and their best-fit linear relation.
Low-S/N$_{\rm line}$ measurements can produce line ratios that severely deviate from their underlying values, particularly in logarithmic space.
In the left panel, which includes low-S/N$_{\rm line}$ galaxies, the log(H$\alpha$/H$\gamma$) distribution is more scattered than log(H$\alpha$/H$\beta$) since H$\gamma$ is the weakest line.
Nevertheless, as clearly seen in Figure~\ref{fig_sample}, the main trend depends little on the S/N$_{\rm line}$ cut.
Moreover, because we employ an error-weighted orthogonal distance regression, the fitting results remain nearly unchanged across the different S/N$_{\rm line}$ cuts.
Note that the error in each line ratio is calculated from the line flux uncertainties via standard error propagation.
Standard error propagation may be unreliable at low S/N$_{\rm line}$, but its impact is negligible in the higher-S/N$_{\rm line}$ samples, so the fitting results for these samples are expected to be robust.
Although we neither force the fit to pass through the intrinsic position nor impose any additional restrictions on the sample galaxies, the resulting relation (orange line) turns out to lie very close to the position (green circle), with $\Delta$log(H$\alpha$/H$\gamma$) = $\sim$0.013 (3\%) at log(H$\alpha$/H$\beta$) = 0.456.
This reflects that our assumption of star-forming galaxies mostly powered by \ion{H}{ii} regions under similar physical conditions is reasonable.

To assess whether the data points are consistent with a single linear relation, we compare the observed orthogonal distances from the best-fit line with those expected from measurement errors. 
The orthogonal distance expected from measurement errors is computed for individual galaxies, accounting for the fitted slope and the covariance term from H$\alpha$ appearing in both line ratios. 
In the S/N$_{\rm line} >$ 1, 5, and 10 samples, the root-mean-square values of the observed distances are 0.065, 0.034, and 0.022 dex, respectively, while the corresponding values expected from measurement errors are 0.063, 0.030, and 0.019 dex.
Given that the observed scatter is dominated by measurement errors, we suggest that the intrinsic variation of nebular attenuation curves is marginal, at least in the wavelength range from H$\gamma$ to H$\alpha$.
We also examine the orthogonal residuals along the best-fit line and find no evidence of systematic curvature.
With increasing S/N$_{\rm line}$, the residual slope diminishes and the scatter becomes more uniform, supporting the validity of a linear fit.

\begin{figure*}[htbp]
\centering
\includegraphics[width=140mm]{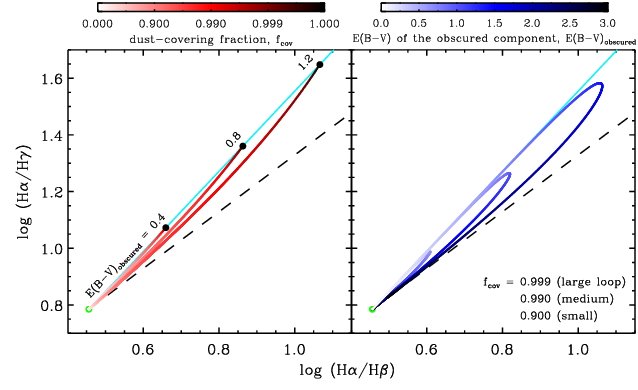}
\caption{Effect of dust-covering fraction on the H$\alpha$/H$\beta$ versus H$\alpha$/H$\gamma$ line ratio diagram. A simplified two-component model is adopted in which the line-emitting regions of a galaxy are divided into dust-obscured (following the Calzetti attenuation curve; cyan line) and unobscured (located at the reference point; green circle) components. In the left panel, the three curves illustrate how the line ratios vary with the dust-covering fraction between the unobscured and obscured components, for $E(B-V)$ values of 0.4, 0.8, and 1.2 (black circles). In the right panel, the three trajectories demonstrate their variation with increasing $E(B-V)$ of the obscured component, for dust-covering fractions of 0.9, 0.99, and 0.999. Note that the dashed line represents the extreme case of a unity slope (i.e., $k_{\rm H\alpha} < k_{\rm H\beta} = k_{\rm H\gamma}$).\label{fig_cover}}
\end{figure*}

Importantly, the observed slope (approximately 1.34, varying by only 0.008 among the S/N$_{\rm line} >$ 1, 5, and 10 samples) is significantly lower than those of the stellar attenuation curves (1.40--1.53), indicating that the average nebular attenuation curve is shallower than typical stellar attenuation curves in the optical.
Note that the uncertainties in our slope and intercept are only 0.002--0.003 and 0.001--0.002, respectively, estimated by bootstrap resampling.
Considering that the stellar attenuation curves are on average steeper than the Calzetti curve in \citet{sal18}, the contrast between nebular and stellar attenuation curves is even more pronounced.
When the galaxies are divided into three bins according to the stellar H$\gamma$ absorption equivalent width ($\rm EW_{H\gamma}$) from the MPA–JHU catalog, the average nebular attenuation curve slopes vary only weakly across the bins (a total variation of 0.010). 
We thus find no evidence that potential inaccuracies in the stellar absorption correction are the primary driver of the observed shallow nebular attenuation curve.
Taken together, nebular attenuation curve slopes change by $\sim$0.01 in each of the systematic tests based on redshift, S/N$_{\rm line}$, and $\rm EW_{H\gamma}$. 
Even if these effects were assumed to accumulate, their combined magnitude would still be insufficient to account for the $\gtrsim$0.06 difference between our slopes and those of commonly adopted stellar attenuation curves.
Therefore, our main conclusion that the nebular attenuation curves are on average shallower than the stellar attenuation curves remains valid.

Among the studies on Balmer line ratios using large samples of nearby galaxies, relatively shallow nebular attenuation curves are also seen in \citet{lin24}, whereas relatively steep nebular attenuation curves are presented in \citet{rez21}. 
Although \citet{rez21} and \citet{lin24} argued that nebular attenuation curves are in agreement with stellar attenuation curves within the uncertainties, these trends can be inferred from their figures.
\citet{rez21} constructed luminosity-averaged composite spectra of SDSS galaxies in several parameter bins and measured each Balmer line using two Gaussian functions representing nebular emission and stellar absorption, without modeling the stellar continuum. 
If the stellar absorption in H$\gamma$ is underestimated relative to H$\alpha$ and H$\beta$, it ultimately leads to steep nebular attenuation curves. 
However, inspecting in detail why \citet{rez21} obtained the opposite result is beyond the scope of this study.
Our slope is comparable to the values calculated from the $k_{\lambda}$ parameterizations in recent studies that jointly analyze Balmer and Paschen lines, although these studies are based on a limited number of distant galaxies: 1.269 for \citet{san25} and 1.323 for \citet{red26a}.

Attenuation curves have often been approximated by a simple power law ($k_{\lambda} \propto \lambda^{-n}$), particularly over narrow wavelength ranges \citep[e.g.,][]{cha00,fit09,cul17,son26}.
For a power-law attenuation curve, the slope in the Balmer line ratio plane converges to 1.378 as the index $n$ approaches zero.
Because the slope we derive is below the theoretical limit, a simple power-law model fails to reproduce the average nebular attenuation curve.
This suggests that $k_{\rm H\gamma}$ is smaller than that expected for a nebular attenuation curve following a simple power-law form.
\citet{lu26} also reported a similar deficit at blue optical wavelengths in stellar attenuation curves.
We note that the gap between the observed slope and the power-law limit is only $\sim$0.04, so the $k_{\rm H\gamma}$ deficit may be sensitive to the accumulated systematic effect.

\subsection{Effect of Dust-Covering Fraction}\label{sec:cover}

The geometric effect on stellar attenuation curves caused by the different spatial distribution between stars and dust has been extensively investigated. 
There is now a general consensus that complex star-dust geometries tend to make stellar attenuation curves shallower, supported by both theoretical \citep[e.g.,][]{wit96,seo16,nar18} and observational \citep[e.g.,][]{wid11,bar20,fis25} methods.
This is primarily because photons that escape without passing through the dust screen yield a lower net optical depth at short wavelengths.
Similar geometric effects also seem to be present in nebular attenuation curves, as discussed in \citet{pre22}, \citet{red26a}, and \citet{woz26}.
By introducing a dust-covering fraction factor, \citet{red26a} successfully resolved the discrepancy between nebular attenuation inferred from Balmer and Paschen line ratios under a given stellar attenuation curve, implying shallower nebular attenuation curves.

Figure~\ref{fig_cover} illustrates the effect of the dust-covering fraction on the Balmer line ratio diagram. 
The inhomogeneous spatial distribution between nebular gas and dust is simplified using a two-component model, in which a fraction of line-emitting regions experiences uniform dust obscuration, while the remainder is unobscured.
When the dust-covering fraction is zero, star-forming galaxies occupy a single point corresponding to the intrinsic line ratios; when the fraction is unity, the line ratios lie along a straight line whose slope is set by the adopted attenuation curve.
For realistic cases where it is between zero and unity, the combined line ratios fall below the straight line.
The left panel shows how the combined line ratios vary with the dust-covering fraction at fixed color excesses for the dust-obscured component.
Conversely, the right panel displays how the combined line ratios vary as a function of the color excess at fixed dust-covering fractions.
As the color excess increases, they initially follow the straight line from the reference point, then bend downward and return toward the reference point because the contribution from the obscured component becomes negligible at sufficiently large optical depths.
A larger dust-covering fraction produces a more extended loop-like trajectory.
Since $k_{\rm H\alpha}<k_{\rm H\beta}<k_{\rm H\gamma}$ holds throughout this toy model, they cannot be located in the region with slopes smaller than unity.
Although we assume the \citet{cal00} attenuation curve in the plot, the same patterns are obtained with other attenuation curves.

\begin{figure*}[!ht]
\centering
\includegraphics[width=\textwidth]{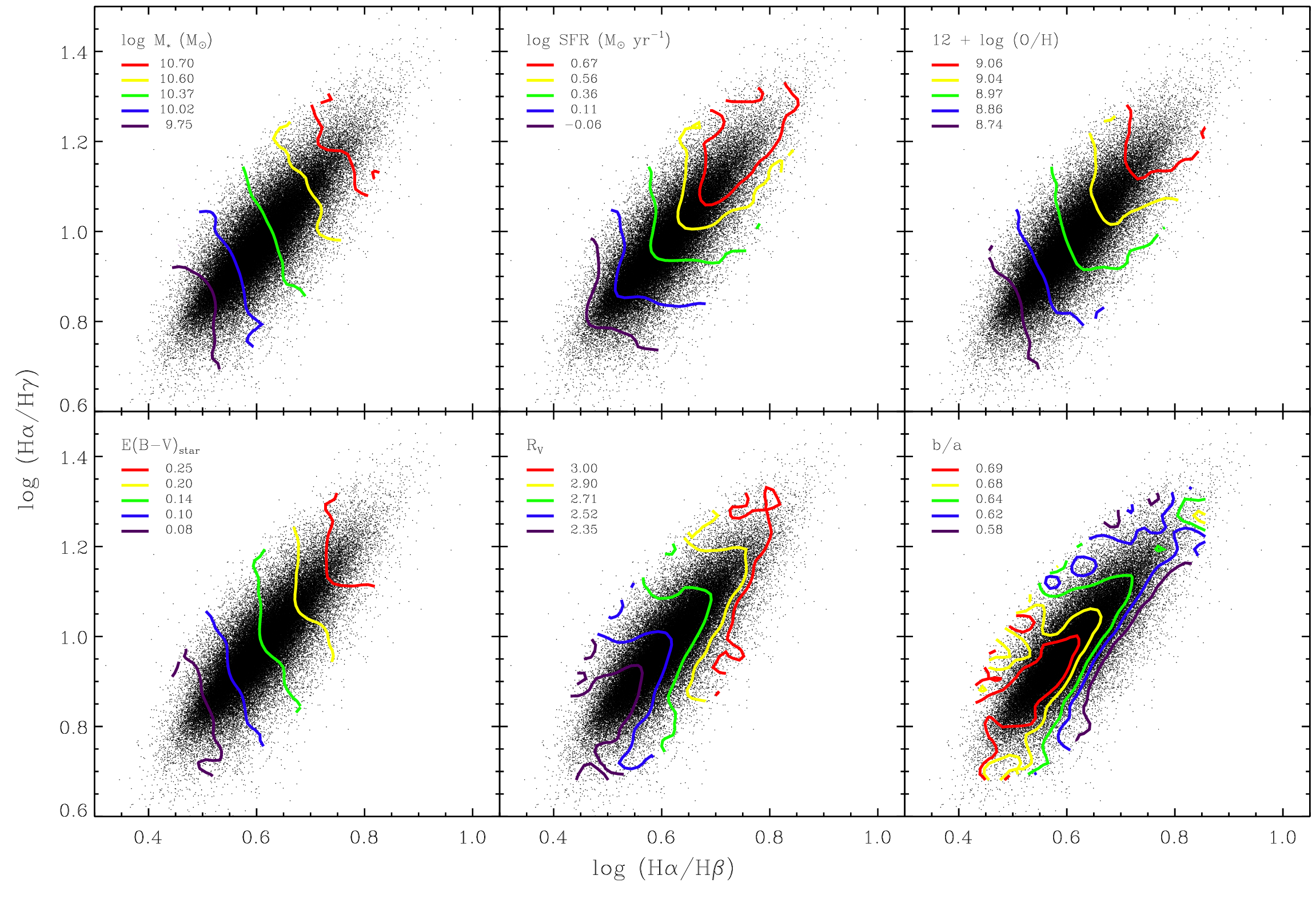}
\caption{H$\alpha$/H$\beta$ versus H$\alpha$/H$\gamma$ line ratio diagram of galaxies with S/N$_{\rm line} >$ 5, 
with contours indicating the levels of six physical parameters: stellar mass (top-left), star formation rate (top-middle), oxygen abundance (top-right), $E(B-V)_{\rm star}$ (bottom-left), $R_V$ (bottom-middle), and axis ratio (bottom-right).\label{fig_contour}}
\end{figure*}

This simple model is intended as a qualitative example of how a non-uniform dust screen can produce effective attenuation curves shallower than those expected from a uniform dust screen. 
We obtain similar behavior with a screen–mixed geometry model (see Appendix~\ref{sec:appen} for more details), indicating that the optical Balmer line ratios alone do not uniquely constrain the underlying source-dust geometry. 
Further studies are needed to investigate more realistic geometric effects.
Spatially resolved spectroscopic data would be useful for directly examining whether the spatial inhomogeneity between dust and nebular gas is more pronounced than that between dust and stellar populations, as a possible explanation for the relatively shallow nebular attenuation curve.
Because our data are based on integrated light, we instead consider the geometric effect indirectly using the galaxy axis ratio (see Section~\ref{sec:depend}). 
A lower axis ratio corresponds to a higher inclination angle, implying not only a higher dust column density but also a more complex geometry \citep[e.g.,][]{che13,bat17}.

\subsection{Dependence on Physical Parameters}\label{sec:depend}

\begin{figure*}[ht!]
\centering
\includegraphics[width=\textwidth]{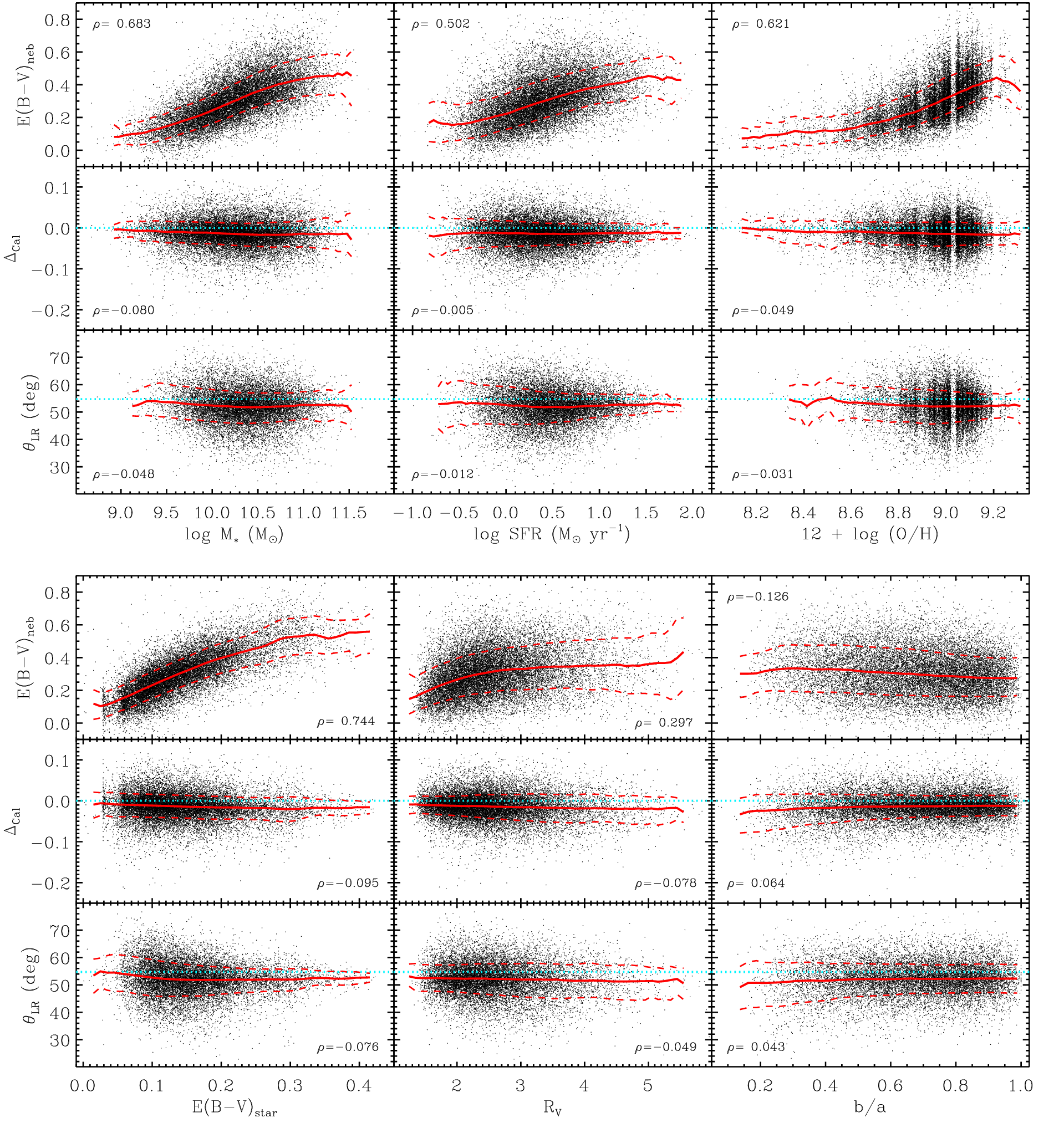}
\caption{$E(B-V)_{\rm neb}$, $\Delta_{\rm Cal}$, and $\theta_{\rm LR}$ as functions of six physical parameters. In the top row panels, they are plotted against stellar mass, star formation rate, and oxygen abundance. In the bottom row panels, they are plotted against $E(B-V)_{\rm star}$, $R_V$, and axis ratio. In each panel, the red solid line represents the sliding median, while the red dashed lines indicate the 16th–-84th percentile range. The Spearman correlation coefficient $\rho$ is denoted in the corner. For clarity, only 20\% of galaxies, randomly selected, are displayed. The \citet{cal00} law is overplotted as a cyan dotted horizontal line in the $\Delta_{\rm Cal}$ and $\theta_{\rm LR}$ panels for comparison.\label{fig_median}}
\end{figure*}

Figure~\ref{fig_contour} shows how six parameters ($M_*$, SFR, 12 + log(O/H), $E(B-V)_{\rm star}$, $R_V$, and $b/a$) of galaxies vary in the H$\alpha$/H$\beta$--H$\alpha$/H$\gamma$ plane.
The overlaid contours are based on mean parameter values computed using galaxies weighted by a spline kernel to obtain smooth distributions.
They are roughly perpendicular to the observed sequence of galaxies rather than parallel to it.
The stellar mass, SFR, oxygen abundance, $E(B-V)_{\rm star}$, and $R_V$ appear to increase with increasing nebular attenuation, while the axis ratio tends to decrease.
However, no parameter exhibits a clear one-directional trend with the nebular attenuation curve slope.

To quantitatively investigate the relations between nebular attenuation and other physical properties, we define three parameters that characterize nebular attenuation.
The nebular color excess is derived as  
  \begin{equation}
  E(B-V)_{\rm neb} = \frac{2.5}{k_{\rm H\beta}-k_{\rm H\alpha}}~\log \left[\frac{({\rm H\alpha}/{\rm H\beta})}{2.857}\right],
  \end{equation}
  where $k_{\rm H\alpha}$ and $k_{\rm H\beta}$ are from the \citet{cal00} attenuation curve.
The trends between $E(B-V)_{\rm neb}$ and other parameters remain unchanged even if $E(B-V)_{\rm neb}$ is derived from H$\alpha$/H$\gamma$ rather than H$\alpha$/H$\beta$. 
We note that the absolute value of $E(B-V)_{\rm neb}$ cannot be determined from this study alone without assuming a functional form for the nebular attenuation curve.
Using the line ratios, the nebular attenuation curve slope is expressed as
  \begin{equation}
  \theta_{\rm LR} = \arctan \left( \frac{\log \left[(\mathrm{H}\alpha/\mathrm{H}\gamma)/6.092\right]}{\log \left[(\mathrm{H}\alpha/\mathrm{H}\beta)/2.857\right]}\right).
  \end{equation}
Because $\theta_{\rm LR}$ is highly uncertain for galaxies with low nebular attenuation, we consider $\theta_{\rm LR}$ to be meaningful only for galaxies with log(H$\alpha$/H$\beta) >$ 0.532 and log(H$\alpha$/H$\gamma) >$ 0.893, corresponding to $E(B-V)_{\rm neb} >$ 0.15, which is satisfied by 78\% of the primary sample.
As an additional parameter related to the attenuation curve shape, $\Delta_{\rm Cal}$ denotes the perpendicular offset from the Calzetti law in the line ratio plane, which is particularly useful for less dusty galaxies.
Larger (smaller) $\theta_{\rm LR}$ indicates a steeper (shallower) nebular attenuation curve,
and positive (negative) $\Delta_{\rm Cal}$ indicates a steeper (shallower) curve relative to the Calzetti law.

\begin{table*}[htbp]
\caption{Partial Spearman correlation coefficients between nebular attenuation and other physical parameters\label{tab_corr}}
\centering
\begin{tabular}{crrrrrr}
\toprule
& \multicolumn{1}{c}{$M_*$} & \multicolumn{1}{c}{SFR} & \multicolumn{1}{c}{12 + log (O/H)} & \multicolumn{1}{c}{$b/a$} & \multicolumn{1}{c}{$E(B-V)_{\rm star}$} & \multicolumn{1}{c}{$R_V$} \\
\midrule
$E(B-V)_{\rm neb}$ & 0.258$^{+0.004}_{-0.004}$ & 0.078$^{+0.004}_{-0.004}$ & 0.346$^{+0.003}_{-0.003}$ & $-$0.373$^{+0.003}_{-0.003}$ & 0.554$^{+0.003}_{-0.003}$ & 0.046$^{+0.004}_{-0.004}$ \\
$\Delta_{\rm Cal}$ & $-$0.065$^{+0.004}_{-0.003}$ & 0.049$^{+0.003}_{-0.004}$ & 0.002$^{+0.003}_{-0.003}$ & 0.050$^{+0.004}_{-0.003}$ & $-$0.078$^{+0.003}_{-0.004}$ & $-$0.025$^{+0.003}_{-0.003}$ \\
$\theta_{\rm LR}$ & $-$0.034$^{+0.004}_{-0.004}$ & 0.018$^{+0.004}_{-0.004}$ & $-$0.006$^{+0.004}_{-0.004}$ & 0.045$^{+0.004}_{-0.004}$ & $-$0.060$^{+0.004}_{-0.004}$ & $-$0.022$^{+0.004}_{-0.004}$ \\
\bottomrule
\end{tabular}
\tabnote{Notes, the quoted upper and lower uncertainties correspond to the 68\% bootstrap confidence interval (i.e., 16th–-84th percentiles). They are nearly identical to the dispersions of the null distributions obtained by randomly shuffling the nebular attenuation parameter for each correlation.}
\end{table*}

Figure~\ref{fig_median} displays the nebular attenuation parameters as functions of stellar mass, SFR, oxygen abundance, $E(B-V)_{\rm star}$, $R_V$, and axis ratio.
The median trends and Spearman rank correlation coefficients are also presented.
As expected in Figure~\ref{fig_contour}, $E(B-V)_{\rm neb}$ is strongly correlated with the six parameters, but $\Delta_{\rm Cal}$ and $\theta_{\rm LR}$ show weaker correlations. 
The strongest correlation is found between $E(B-V)_{\rm neb}$ and $E(B-V)_{\rm star}$, and we confirm that $E(B-V)_{\rm neb}$ is a factor of two larger than $E(B-V)_{\rm star}$, compatible with a scenario in which nebular gas and young stars are embedded in dusty clouds, while older stars are mixed with the diffuse interstellar medium \citep[e.g.,][]{cal94,koy19,tsu26}.
The nebular-to-stellar reddening ratio in this work depends on the adopted attenuation curve used to derive $E(B-V)_{\rm neb}$. 
The Calzetti curve yields the smallest $E(B-V)_{\rm neb}$ value among the five typical attenuation curves, but the difference is about 15.6\% relative to the largest value derived using the Cardelli curve.
Although the correlations of $\theta_{\rm LR}$ and $\Delta_{\rm Cal}$ with $R_V$ are not strong, we suggest that galaxies with shallower (steeper) stellar attenuation curves tend to have shallower (steeper) nebular attenuation curves.
The strong correlation between the amounts of nebular and stellar attenuation supports a physical connection between them, while the weak correlation between their attenuation curve slopes is consistent with such a connection.

The six parameters considered in comparison with the nebular attenuation parameters are interdependent.
The fundamental relation among galaxy mass, SFR, and metallicity is well established \citep[e.g.,][]{yat12,bot13,kor25}.
Stellar attenuation properties are also closely linked to other galaxy properties, including stellar mass, SFR, metallicity, and inclination \citep[e.g.,][]{zah17,sal18,lu26}.
Thus, we compute partial Spearman correlation coefficients to assess the intrinsic relationship between nebular attenuation and other parameters by controlling for $M_*$, SFR, 12 + log(O/H), and $b/a$.
Naturally, the variables of interest are excluded from the control set in each case.
We do not include $E(B-V)_{\rm star}$ and $R_V$ as control variables to avoid over-controlling, as they are physically coupled to the nebular attenuation parameters and would remove part of the intrinsic signal we aim to measure.
In analyzing the correlations of $E(B-V)_{\rm neb}$ with other parameters, $E(B-V)_{\rm star}$ should be excluded from the control set because it acts as a mediator, whereas including or excluding $R_V$ has little impact on the correlations.
Conversely, for the correlations of $\Delta_{\rm Cal}$ and $\theta_{\rm LR}$ with other parameters, $R_V$ should be excluded from the control set, whereas including or excluding $E(B-V)_{\rm star}$ has little impact.
Specifically, when computing the correlation between $E(B-V)_{\rm neb}$ and $M_*$, we control for SFR, 12 + log(O/H), and $b/a$.
For the correlation between $E(B-V)_{\rm neb}$ and $E(B-V)_{\rm star}$, we control for $M_*$, SFR, 12 + log(O/H), and $b/a$.
We then estimate bootstrap confidence intervals and perform a noise-only null test to evaluate the significance of the correlations.
These results are presented in Table~\ref{tab_corr}.
Although the correlations exhibit a range of strengths, all correlation coefficients are significantly different from zero, except for the correlations of $\Delta_{\rm Cal}$ and $\theta_{\rm LR}$ with 12 + log(O/H).
Notably, the $\Delta_{\rm Cal}$--SFR and $\theta_{\rm LR}$--SFR trends reverse sign and become slightly stronger in the partial correlations.

Similar to stellar attenuation, nebular attenuation is known to show positive correlations with stellar mass, SFR, metallicity, and inclination \citep[e.g.,][]{zah17,li21,lor24,lee25}.
These relations are also clearly seen in our $E(B-V)_{\rm neb}$ results in Table~\ref{tab_corr}. 
We again note that the axis ratio $b/a$ is anti-correlated with galaxy inclination.
The dependence of stellar attenuation curve slopes on physical parameters has been reported to be diverse.
For example, with increasing stellar mass, some studies find a flattening of the slope \citep[e.g.,][]{bat16,sal18,bar20}, but others find no strong trend \citep[e.g.,][]{boq22,fis25} or even a steepening \citep[e.g.,][]{zei15,lu26}.
These conflicting results may arise from differences in the samples, wavelength ranges, and methods used to measure the slopes, which reflect different aspects of dust attenuation and make meaningful comparisons difficult.
In particular, \citet{sal18} demonstrated that stellar attenuation curve slopes are mainly driven by dust opacity, becoming shallower with increasing $A_V$. 
They suggested that the apparent flattening with increasing stellar mass or inclination is likely a secondary effect that is substantially reduced at fixed $A_V$.
They found no significant dependence on gas-phase metallicity.
However, the dependence of nebular attenuation curve slopes on physical parameters remains largely unexplored.
In this work, we provide the first direct comparison between nebular and stellar attenuation curve slopes.
The anti-correlations of $\Delta_{\rm Cal}$ and $\theta_{\rm LR}$ with $R_{\rm V}$ are weak ($|\rho| =$ 0.02–-0.03) but statistically significant ($>5\sigma$), which are consistent with a positive correlation between the nebular and stellar attenuation curve slopes.
As expected from this connection, the nebular attenuation curve slope appears to broadly follow the physical trends for the stellar attenuation curve slope reported by \citet{sal18}, becoming shallower with increasing opacity, stellar mass, and inclination, while showing little dependence on metallicity.
The inclination dependence of attenuation curve slopes highlights the importance of source–dust geometry in shaping attenuation curves, while their little metallicity dependence implies that dust chemical composition is less important than source-dust geometry.
\citet{sal18} also found that the stellar attenuation curve slope has a non-monotonic dependence on sSFR, with shallow slopes for star-forming main sequence galaxies and steeper slopes toward lower- and higher-sSFR galaxies.
The positive partial correlations of $\Delta_{\rm Cal}$ and $\theta_{\rm LR}$ with SFR can be understood as a consequence of our selection bias toward higher-SFR galaxies at fixed stellar mass due to the requirement of H$\gamma$ detection, as seen in Figure~\ref{fig_sample}.

\section{Summary}\label{sec:summ}

For nearby star-forming galaxies with previously characterized stellar attenuation curves, we constrain their nebular attenuation curves using H$\alpha$, H$\beta$, and H$\gamma$ emission lines.
We then compare the slopes of the nebular attenuation curves with those of the stellar attenuation curves and examine their correlations with other physical parameters. 
Our main findings below are based on the primary sample of 90,958 galaxies with S/N$_{\rm line} >$ 5.

\begin{itemize}
\item In the log(H$\alpha$/H$\beta$) versus log(H$\alpha$/H$\gamma$) diagram, the average nebular attenuation curve (slope = $1.337\pm0.002$) is significantly shallower than typical stellar attenuation curves (slope = 1.40--1.53).
\item The line ratio variations driven by the dust-covering fraction demonstrate that non-uniform source–dust geometries can contribute to the relatively shallow nebular attenuation curve.
\item The nebular and stellar attenuation curve slopes are weakly but positively correlated. Although the effect size is small, galaxies with shallower (steeper) stellar attenuation curves tend to have shallower (steeper) nebular attenuation curves.
\item The nebular attenuation curve slope shows weak dependences on galaxy properties, becoming shallower with increasing stellar mass, SFR, and inclination, but shows little dependence on gas-phase metallicity. These trends broadly follow those of the stellar attenuation curve slope.
\end{itemize}

Our results are insensitive to the S/N$_{\rm line}$ threshold and are thus representative of the general population of nearby star-forming galaxies.
However, as our analysis is limited to a narrow wavelength coverage, it may not fully capture the intrinsic variations in nebular attenuation curves.
In addition, spatially resolved data are required to better understand the geometric effects shaping attenuation curves by comparing stellar and nebular attenuation distributions.


\acknowledgments
We thank the anonymous referee for a careful reading of the manuscript and constructive comments.
This research was supported by the Korea Astronomy and Space Science Institute under the R\&D program (Project No. 2026-1-831-00), supervised by the Korea AeroSpace Administration.
H.Shim was supported by the National Research Foundation of Korea (NRF) grant funded by the Korea government (MSIT) (No. RS-2024-00349364). 
H.S.Hwang acknowledges support from the National Research Foundation of Korea (NRF) funded by the Korea government (MSIT; RS-2026-25482692) and the Global-LAMP Program funded by the Ministry of Education (RS-2023-00301976).

\appendix

\section{Effect of Screen-Mixed Geometry}\label{sec:appen}

\begin{figure}[htbp]
\centering
\includegraphics[width=\columnwidth]{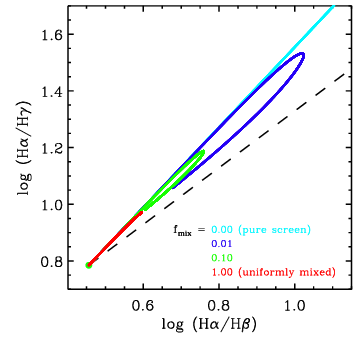}
\caption{Effect of screen-mixed geometry on the H$\alpha$/H$\beta$ versus H$\alpha$/H$\gamma$ line ratio diagram. The intrinsic line ratio position (green circle) lies in the lower-left corner. The Calzetti attenuation curve (cyan line) is adopted as the underlying screen relation. The blue, green, and red trajectories demonstrate the line ratio variations with increasing optical depth for mixed-component fractions of 0.01, 0.10, and 1.00, respectively. For clarity, only the range of $0 < \tau_V \leq 10$ is shown. The dashed line serves as a guide to a slope of unity.\label{fig_appen}}
\end{figure}

In Figure~\ref{fig_appen}, we examine another simple two-component model, which combines a foreground screen and a uniformly mixed geometry (hereafter, screen-mixed geometry model),
  \begin{equation}
  F_\lambda = F_{\lambda,0}\left[(1-f_{\rm mix})e^{-\tau_\lambda}+f_{\rm mix}\frac{1-e^{-\tau_\lambda}}{\tau_\lambda}\right],
  \end{equation}
 where $F_\lambda$ and $F_{\lambda,0}$ are the observed and intrinsic fluxes at wavelength $\lambda$, respectively, $f_{\rm mix}$ is the fraction of the emission arising from the uniformly mixed component, and $\tau_{\lambda}$ is the dust optical depth ($=\tau_V k_\lambda/R_V$). 
In the uniformly mixed case ($f_{\rm mix}$ = 1), the line ratios start from their intrinsic values and asymptotically approach log(H$\alpha$/H$\beta$) = 0.597 and log(H$\alpha$/H$\gamma$) = 0.973 as $\tau_V$ increases. 
For $0 < f_{\rm mix} < 1$, the line ratios initially follow the pure screen relation, but eventually turn back toward the asymptotic values. 
A smaller $f_{\rm mix}$ produces a more extended loop-like trajectory. 
As a result, like the dust-covering fraction model (see Section~\ref{sec:cover}), the screen-mixed geometry model generates effectively shallower attenuation curves than the underlying attenuation curve. 
Based on the optical Balmer line ratios alone, it is challenging to distinguish between these two models.





\end{document}